\documentclass[aps,twocolumn,pra,groupedaddress,amsmath,amssymb]{revtex4-2}
\usepackage{dcolumn}
\usepackage{bm}
\usepackage{amsmath}
\usepackage{txfonts}
\usepackage[T1]{fontenc}
\usepackage{xspace}
\usepackage{ulem}
\usepackage{comment}
\usepackage{braket}
\ifx\pdfoutput\undefined
\usepackage[dvipdfmx]{graphicx}
\usepackage[dvipdfmx]{hyperref}
\usepackage[dvipdfmx]{color}
\else
\usepackage{graphicx}
\usepackage{hyperref}
\usepackage{color}
\fi

\begin{document}
\let\emph\textit

\newcommand{\cred}[1]{\textcolor{black}{#1}}
\newcommand{\cblue}[1]{\textcolor{blue}{#1}}

\title{Measurement-induced phase transitions for free fermions in a quasiperiodic potential}
\author{Toranosuke Matsubara}
\email{matsubara.t.6125@m.isct.ac.jp}
\author{Kazuki Yamamoto}
\author{Akihisa Koga}

\affiliation{
  Department of Physics, Institute of Science Tokyo,
  Meguro, Tokyo 152-8551, Japan
}
\altaffiliation{Former Tokyo Institute of Technology
}

\date{\today}
\begin{abstract}
We study the dynamics under continuous measurements for free fermions in a quasiperiodic potential by
using the Aubry-Andr\'{e}-Harper model with hopping rate $J$ and potential strength $V$. On the basis of the quantum trajectory method, we obtain the phase diagram for the steady-state entanglement entropy and demonstrate that robust logarithmic system-size scaling emerges up to a critical potential strength $V_c/J \sim 2.3$. Moreover, we find that the measurement induces entanglement phase transitions
from the logarithmic-law phase to the area-law phase for the potential strength $V< V_c$, while any finite measurement stabilizes the area-law phase for $V>V_c$. This result is distinct from the entanglement scaling
in the unitary limit,
where volume-law and area-law phases undergo a transition at $V/J=2$.
To further support the phase diagram, 
we analyze the connected correlation function and find that it shows algebraic decay in the logarithmic-law phase, while it decays quickly in the area-law phase. 
Our results can be tested in ultracold atoms by introducing quasiperiodic potentials and continuously monitoring the local occupation number with an off-resonant probe light.
\end{abstract}
\maketitle

\section{Introduction}
Entanglement entropy is a fundamental quantity in quantum mechanics
and is of central interest
in the exploration of modern quantum technologies and materials~\cite{Amico_2008, Horodecki_2009, Eisert_2010, Laflorencie_2016}.
It also underpins the advance in quantum computation~\cite{Ladd_2010, Albash_2018}
and quantum information processing~\cite{Vedral_2002, Braunstein_2005, Weedbrook_2012}.
Recently, the research of entanglement entropy has been broadened to open quantum systems, where interactions with environments are inevitable, particularly to the topic of the effects of measurements. One of the fascinating phenomena is the measurement-induced phase transition (MIPT), where measurements drastically change the scaling behavior of the entanglement entropy and induce transitions between distinct entanglement phases \cite{Li_2018, Chan_2019, Skinner_2019}.
These transitions reflect the competition between unitary evolution governed by the system Hamiltonian and quantum jumps induced by measurements.
Over the past few years, MIPTs have been extensively studied, {e.g., in random quantum circuits under measurements~\cite{Szyniszewski_2019, Li_2019, Bao_2020, Choi_2020, Gullans_2020a, Gullans_2020b, Jian_2020, Zabalo_2020, Iaconis_2020, Turkeshi_2020, Zhang_2020, Szyniszewski_2020, Nahum_2021, Ippoliti_2021a, Ippoliti_2021b, Lavasani_2021a, Lavasani_2021b, Sang_2021, Lu_2021, Fisher_2022, Block_2022, Sharma_2022, Agrawal_2022, Barratt_2022, Kelly_2023, Shkolnik_2023, Zabalo_2023} and in many-body trajectory dynamics under continuous monitoring~\cite{Tang_2020, Goto_2020, Fuji_2020, Doggen_2022, Doggen_2023, Lunt_2020, Yamamoto_2023, De_2024, Patrick_2024}.
Experimentally, MIPTs have been observed in superconducting qubits~\cite{Koh_2023, Google_2023} and trapped-ion systems~\cite{Noel_2022, Agrawal_2024}, which lend credence to the theoretical predictions.

It is worth noting that the MIPT in free-particle systems is one of the most actively investigated subjects~\cite{Cao_2019, Alberton_2021, Chen_2020, Tang_2021, Coppola_2022, Ladewig_2022, Carollo_2022, Yang_2022, Buchhold_2021, VanRegemortel_2021, Youenn_2023, Loio_2023, Turkeshi_2022, Kells_2023, Yu_2025,Fava_2023,Piccitto_2022,Piccitto_2023,Russomanno_2023}.
Recently,
the absence of MIPTs in free fermions has been reported in one-dimensional (1D) tight-binding models on the spatially homogeneous systems by analyzing the nonlinear sigma models as an effective field theory \cite{Poboiko_2023, Fava_2024,Starchl_2025}.
On the other hand, the existence of MIPTs has been clarified in free fermions above one-dimension \cite{Chahine_2024, Poboiko_2024, Jin_2024}, in free fermions and bosons with long-range couplings \cite{Minato_2022, Muller_2022, Yokomizo_2024}, and in disordered free fermions \cite{Szyniszewski_2023, Popperl_2023}. Therefore, many open questions remain about the universality and mechanisms underlying MIPTs in free fermions with yet unnoticed structures.

As most of the existing studies on MIPTs have considered
homogeneous and disordered systems,
it is timely to focus on another system with incommensurate structures, or quasiperiodic systems, which are characterized by aperiodic but long-range ordered structures.
In quantum circuits, several studies have analyzed such problems, where the measurement profiles are modulated to follow the quasiperiodicity with respect to sites~\cite{Shkolnik_2023}, strengths~\cite{Zabalo_2023}, and spatiotemporal distributions~\cite{Li_2019, Lu_2021}.
\cred{In the context of non-Hermitian quasiperiodic systems, several properties of entanglement entropy have been also studied~\cite{Zhou_2024,Li_2024}.}
Thus, quasiperiodic systems under random measurement offer a rich possibility for studying MIPTs.

However,
dynamics under continuous measurement, where quasiperiodicity is imposed on parameters in the system Hamiltonian such as potentials and lattice configurations, remains largely unexplored.
This setup is one of the most fundamental platforms in condensed matter physics, and it is now possible to control quasiperiodic structures
in optical lattices by using ultracold atomic gases, such as in 1D~\cite{Fallani_2007, Roati_2008, DErrico_2014, Schreiber_2015, Bordia_2017, An_2021, Rajagopal_2019,Yao_2019,Yao_2020}, pentagonal~\cite{Guidoni_1997, Guidoni_1999, Corcovilos_2019,Palencia_2005}, and octagonal~\cite{Viebahn_2019, Sbroscia_2020, Yu_2024,Gautier_2021,Zhu_2023} systems. Experimentally,} these structures are created by superimposing several standing waves with incommensurate wavelengths or angles.
Importantly, quasiperiodic systems exhibit unique localization phenomena distinct from both
homogeneous and disordered systems.
For example, the Aubry-Andr\'{e}-Harper (AAH) model~\cite{Harper_1955, Aubry_1980}, which describes free fermions in a 1D lattice under a quasiperiodic potential, displays transitions
in all eigenstate wave functions between extended and localized states with increasing the potential strength~\cite{Jagannathan_2021}.
These localization properties also manifest in the scaling behavior of steady-state entanglement entropy as
the volume law and area law, respectively~\cite{Roosz_2014, Roy_2021}. Therefore, we ask for the following question: Are there MIPTs in continuously monitored free fermions under quasiperiodic potentials and how does the measurement affect the entanglement scaling?

In this paper, we study MIPTs for free fermions
in the quasiperiodic potential by
using the AAH model with hopping rate $J$ and potential strength $V$.
We analyze the dynamics of entanglement entropy under continuous measurement of the local occupation number and identify MIPTs by focusing on the scaling behavior of the entanglement entropy.
In contrast to the unitary limit, we find a robust logarithmic scaling up to a critical potential strength $V_c/J \sim 2.3$ and demonstrate that the measurement induces MIPTs from the logarithmic- to area-law phase.
On the other hand, we find that any finite measurement immediately leads to the area-law phase for $V>V_c$. Moreover, by employing the finite-size scaling analysis of the entanglement entropy,
we find consistency with the Berezinskii-Kosterlitz-Thouless (BKT) universality
and obtain an entanglement phase diagram.
To further support the phase diagram, we analyze the scaling behavior of correlation functions, which show algebraic decay in the logarithmic-law phase, while it decays quickly in the area-law phase.
Finally, we also show an agreement with the behavior of
the amplitudes of single-particle wave functions; they tend to a constant at the edges in the logarithmic-law phase, but show a localization in the area-law phase.

The rest of this paper is organized as follows.
First, in Sec.~\ref{sec: model}, we introduce the AAH model and outline the quantum trajectory method for simulating the continuously monitored dynamics.
Then, in Sec.~\ref{sec: unitary}, we present a brief summary of the unitary limit.
Sec.~\ref{sec: EE} is devoted to the entanglement phase diagram for MIPTs, and we analyze the measurement-induced properties of entanglement entropy. In Sec.~\ref{sec: support}, we study the connected correlation functions,  single-particle wave functions, and autocorrelation functions to further support the phase diagram. Finally, the summary is given in Sec.~\ref{sec: summary}.

\section{Model and method}\label{sec: model}

We consider free fermions in a 1D lattice under a quasiperiodic potential. The system is described by the AAH model~\cite{Harper_1955, Aubry_1980}
\begin{eqnarray}
  H =-J \sum_{j} \left[  c_j^{\dagger} c_{j+1} + {\rm H.c.}\right]
  +V\sum_{j} \cos \left(2\pi j/\tau + \theta \right) n_j,
  \label{eq: Hamiltonian}
\end{eqnarray}
where $ c_j $ ($ c_j^\dagger $) annihilates (creates) a fermion at the $ j $th site, $ n_j = c_j^\dagger c_j $ denotes the particle-number operator, and $ \tau = (\sqrt{5} + 1)/2 $ is the golden mean.
The parameters $ J $, $ V $, and $ \theta $ stand for the hopping rate to the neighboring sites, the potential strength, and the phase shift of the potential, respectively.
Since $\tau$ is irrational, there exists no periodicity in the system,
and we can discuss the effect of the quasiperiodicity on a free fermion system.
In the following, we consider the AAH model under open boundary condition.

In the study, we simulate continuously monitored dynamics of free fermions 
in the quasiperiodic potential.
We employ the quantum trajectory method,
which is based on the stochastic Schr\"{o}dinger equation.
Under the measurement of the local particle number $ n_j $,
the evolution of the quantum state within a time interval $[t, t+dt]$ is given by
\begin{equation}
  d\left|\Psi\left\{\xi_{j, t}\right\}\right\rangle = \left[-i H \, dt + \sum_{j} \xi_{j, t}\left(\frac{n_j}{\sqrt{\left\langle n_j\right\rangle_t}}-1\right)\right] \left|\Psi\left\{\xi_{j, t}\right\}\right\rangle,
  \label{eq: QJ}
\end{equation}
where $\langle\cdots\rangle_t$ stands for an expectation value with
the quantum state $|\Psi\rangle$. 
$\xi_{j,t}$ is a discrete random variable that is chosen according to $ \xi_{j, t}^2 = \xi_{j, t} $ and $ \overline{\xi_{j, t}} = \gamma \braket{n_j}_t dt$~\cite{Daley_2014, Dalibard_1992, Wiseman_1993, Fuji_2020, Yamamoto_2025}. When a jump occurs at the $j$th site at time $t$, 
$\xi_{j,t} = 1 $, and $ \xi_{j,t} = 0 $ otherwise. Here, $\gamma$ is the measurement strength, and $\overline X$ represents an ensemble average of $X$ over the stochastic process.
In this calculation, the quantum state $|\Psi\rangle$ is represented by the Gaussian states. 
Details of the numerical simulations are provided in Appendix~\ref{app: method} and 
useful numerical formulas for physical quantities are outlined in Appendix~\ref{app: obs}.

Specifically, the ensemble average of Eq.~\eqref{eq: QJ} reduces to the Lindblad master equation, which describes the Markovian dynamics of the density matrix. This equation is commonly used to simulate open quantum systems, particularly in atomic, molecular, and optical physics~\cite{Diehl_2008, Muller_2012, Daley_2014, Yamamoto_2021}.
Both trajectory dynamics and Lindblad dynamics lead to identical results
for the linear quantities with respect to the density matrix.
However, these two descriptions generally yield different results for nonlinear observables. 
Here, we consider the entanglement entropy as one of the representative nonlinear quantities 
and obtain MIPTs.
In our study, we calculate the von Neumann entanglement entropy, which
is defined as
\begin{equation}
S = -\operatorname{tr}_A\left({\rho_{A} \log \rho_{A}}\right),
\end{equation}
where $ \rho_{A} $ is the reduced density matrix of subsystem $A$.
We set the subsystem to ${A} = [1, L/2]$,
where $L$ is the system size.
In our simulations, we set an initial state to the N\'eel state, whose entanglement entropy is zero.
When time evolves, $S$ increases, and eventually saturates in the long-time limit.
(For the transient behavior before reaching the steady-state, see Appendix~\ref{app: steady}).

We evaluate several quantities in the steady state averaging over stochastic quantum trajectories. In some cases, we take an average over the phase shifts $ \theta $ as well. The phase shift considered here is $\theta = 2\pi n/N_{\rm pot}$ with 
$n\:(=1,2,\cdots, N_{\rm pot})$, where $N_{\rm pot}$ is the number of realizations for the phase shifts.
Therefore, the total number of realizations is given by $N = N_{\rm pot} \times N_{\rm traj}$,
where $N_{\rm traj}$ is the number of trajectories per a single realization of the phase shift.

\section{Entanglement property in the unitary limit}\label{sec: unitary}

First, we explain important properties of the AAH model
in the unitary limit $(\gamma /J=0)$.
Applying a Fourier transformation to the Hamiltonian~\eqref{eq: Hamiltonian},
one obtains the AAH Hamiltonian with the hopping rate $V/2$ and the potential strength $2J$.
This means the presence of a self-duality at $V/J = 2$.  
It is known that all eigenstates are extended (localized)
in real space for $V/J < 2$ ($V/J > 2$),
and critical at $V/J = 2$~\cite{Jagannathan_2021}.

\begin{figure}[htb]
  \begin{center}
    \includegraphics[width=0.8\linewidth]{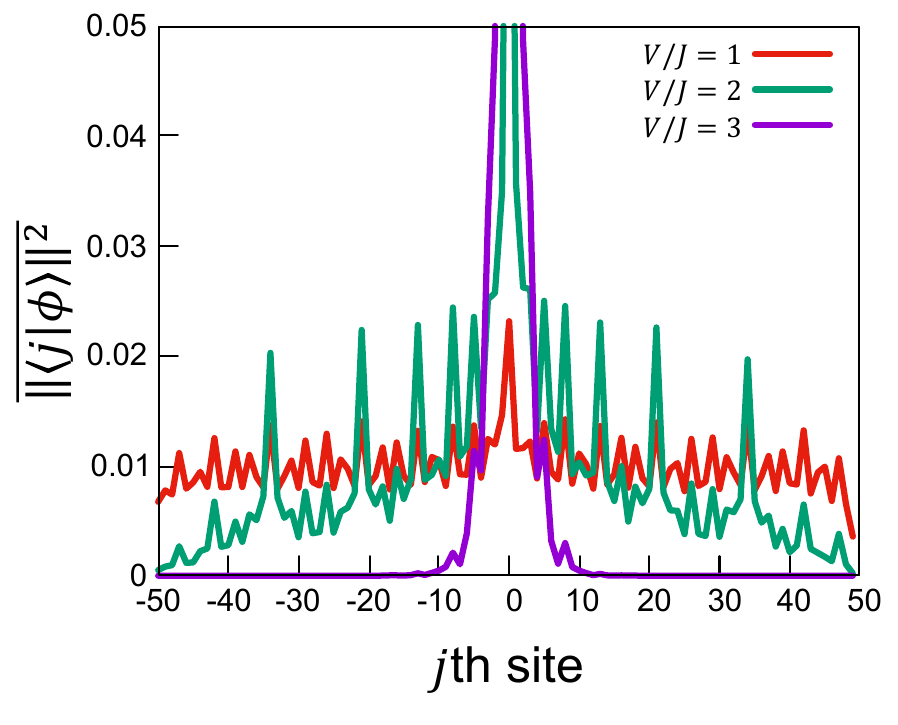}
    \caption{
    Amplitudes of single-particle wave functions in the long-time limit.
    We choose $ V/J = 1 $ (volume-law phase), $ V/J = 2 $ (critical point), and $ V/J = 3 $ (area-law phase), respectively.
    In this simulation, we use $L=100$ and $(N_{\rm pot}, N_{\rm traj}) = (5\times 10^3,1)$.
    }
    \label{fig: unitary_single}
  \end{center}
\end{figure}
This affects the time evolution of any wave functions.
We examine the long-time unitary dynamics of
the N\'eel state with $L/2$ fermions.   
We show in Fig.~\ref{fig: unitary_single} the single-particle wave function in the long time limit,
where its amplitude is maximum at the center of the system.
When $V/J=1$, the amplitude is finite in the whole system
since all eigenstates are extended. 
On the other hand, when $V/J=3$, the wave function is localized around the center of the system
due to the localized eigenstates.
When $V/J=2$, all eigenstates are critical, and
thereby intermediate behavior appears in the wave function.
\begin{figure}[htb]
  \begin{center}
    \includegraphics[width=0.8\linewidth]{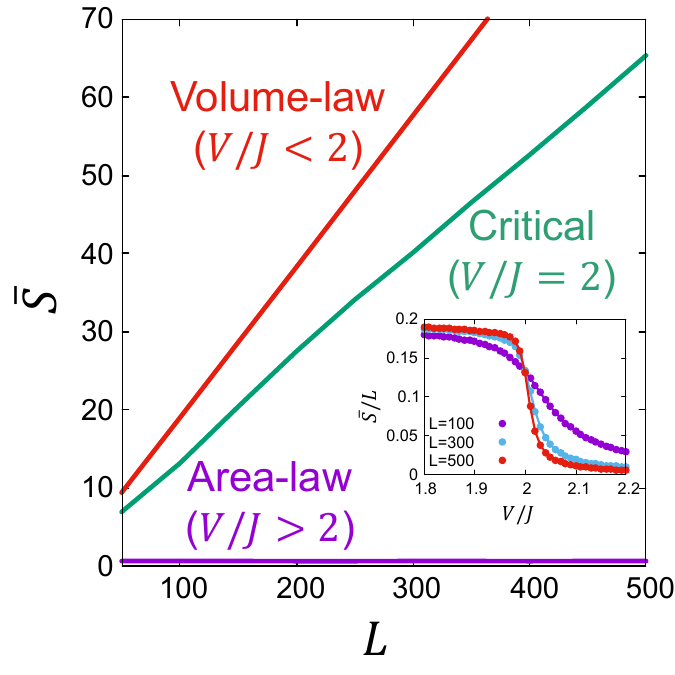}
    \caption{
    System-size dependence of the averaged entanglement entropy $ \overline{S}$.
    The inset shows the potential dependence of slope.
    In this simulation, we use $(N_{\rm pot}, N_{\rm traj}) = (5\times 10^3,1)$.
    }
    \label{fig: unitary_EE}
  \end{center}
\end{figure}
It is also known that 
the entanglement entropy, which is closely related to the wave function, 
is significantly affected by $V/J$~\cite{Roosz_2014, Roy_2021}.
We show in Fig.~\ref{fig: unitary_EE} the system-size dependence of the entanglement entropy.
When $V\le 2J$ ($V>2J$), it follows
the volume law $ \overline{S} \propto L $
(the area law $ \overline{S} \propto L^0 $).
In addition, as shown in the inset of Fig.~\ref{fig: unitary_EE}, 
we find a drastic change in its slope $S/L$ around $V/J=2$.
Notably, the lines for $L=100, 300$, and $500$ intersect at $V/J=2$
and the slope approaches a certain value for $V/J\neq 2$ as $L$ increases.
Therefore, by considering the entanglement entropy,
the system can be classified into two distinct phases.
When $V/J < 2$, the quantum state exhibits volume-law scaling of entanglement entropy,
while for $V/J>2$, it follows area-law scaling.
For $V/J = 2$, the entanglement entropy also follows the volume-law although the slope is smaller than that for $ V/J < 2$.
In the following, these phases are referred to as the volume-law phase and 
area-law phase I, respectively.

\section{Entanglement property under continuous measurement}\label{sec: EE}
\subsection{Phase diagram and steady-state entanglement entropy}

We study the effect of continuous measurement on free fermions in the quasiperiodic potential.
It has been clarified that, in several free-fermionic systems, 
the quantum state with the logarithmic scaling entanglement (logarithmic-law phase)
is induced by the continuous measurements~\cite{Minato_2022, Muller_2022, Szyniszewski_2023}.
The phase should be described by 
\begin{equation}
  \overline{S} = \frac{c_\mathrm{eff}}{3} \ln \left(\frac{L}{\pi} \sin \frac{\pi \ell}{L}\right) + s_0,
  \label{eq: EE}
\end{equation}
where
$c_\mathrm{eff}$ is the effective central charge, $s_0$ is the residual entropy,
and $\ell $ is the subsystem size~\cite{comment}.
Since $c_{\rm eff}$ would become zero in the area-law phase, 
Eq.~\eqref{eq: EE} is important 
to clarify whether or not the logarithmic-law phase is induced by the continuous monitoring
even in the quasiperiodic systems.

\begin{figure}[htb]
  \begin{center}
    \includegraphics[width=\linewidth]{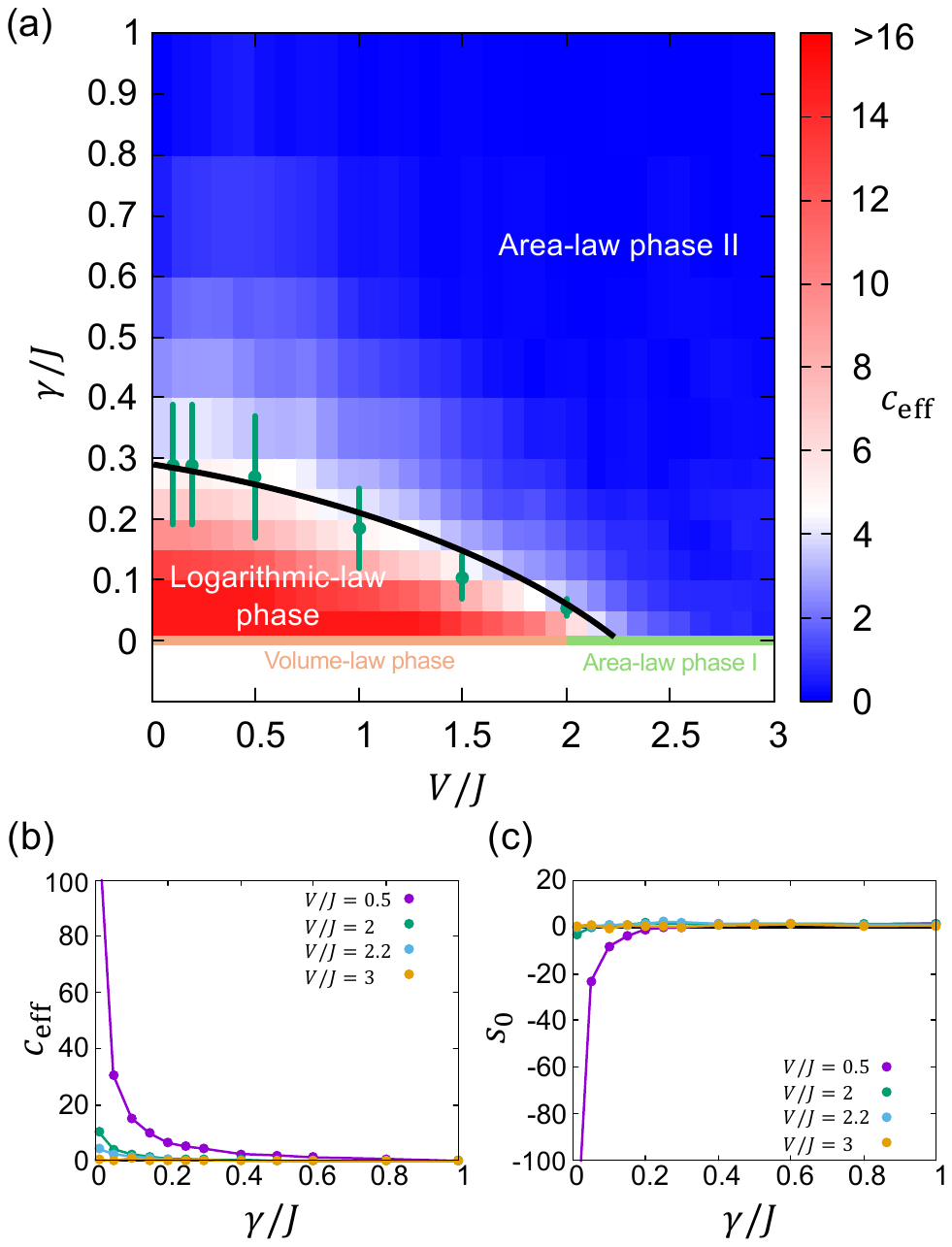}
    \caption{
    (a) Phase diagram for the steady-state entanglement entropy in the AAH model under continuous measurement with a color bar for the effective central charge $c_\mathrm{eff}$, where values above 16 are uniformly displayed with a red color. The black curve represents a guide to the eye,
    where the transition boundary is estimated via the finite-size scaling of the entanglement entropy (see Sec.~\ref{sec: scaling}).    
    The volume-law phase and the area-law phase I in the unitary limit $\gamma /J=0$ are also shown for comparison.
    The fitting is performed by using the data for $L = 300$, 350, 400, 450, and 500.
    (b) [(c)] Effective central charge (residual entanglement entropy) with respect to the measurement strength $\gamma$ for $V/J =$ 0.5, 2, 2.2, and 3.
    In this simulation, we use $(N_{\rm pot}, N_{\rm traj}) = (1,100)$.
    }
    \label{fig: phase-diagram}
  \end{center}
\end{figure}

We conduct a numerical calculation for $\overline{S}$ by following the quantum trajectory method \eqref{eq: QJ} and perform the fitting with the use of Eq.~\eqref{eq: EE}. Then,  
we deduce the effective central charge $c_\mathrm{eff}$ and the residual entanglement entropy $s_0$.
In this calculation, we have confirmed that
100 trajectories are enough to obtain the averaged entanglement entropy 
since the mean error is two orders of magnitude smaller than the averaged value.
We show the profile of $c_{\rm eff}$ in Fig.~\ref{fig: phase-diagram}(a).
It is found that $c_\mathrm{eff}$ is finite
for small values of $\gamma$ and $V$, 
indicating the presence of the logarithmic-law phase.
The area-law phase with $c_{\rm eff}\sim 0$ is realized in the other region. 
Since this measurement-induced phase is distinct from the area-law phase I induced by a strong quasiperiodic potential in the unitary limit,
we refer it as the area-law phase II.
In this analysis, we have numerically confirmed that 
the phase shift little affects the entanglement entropy.
Therefore, in this section,
a single realization of the phase shift ($N_{\rm pot} = 1$)
is adopted to reduce the computational cost.

\begin{figure}[htb]
  \begin{center}
    \includegraphics[width=\linewidth]{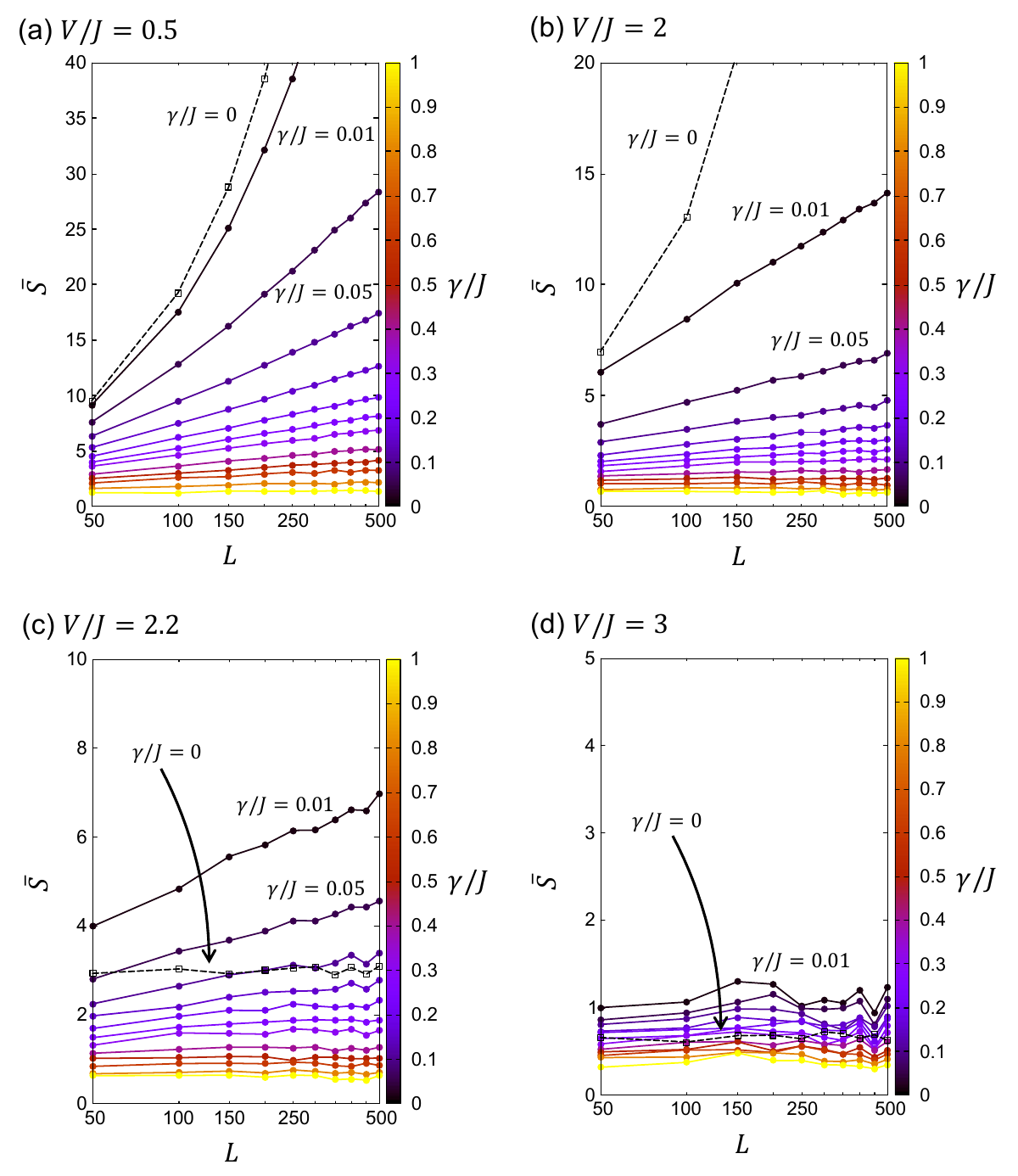}
    \caption{
      System-size dependence of the entanglement entropy for (a) $V/J =1$, (b) $2$, (c) $2.2$, and (d) $3$.
      The horizontal axis is set to be a logarithmic scale.
      The color map represents the values of the measurement strength $\gamma$.
      The entanglement entropy at $\gamma / J = 0$ is depicted by the dashed line.
      In this simulation, we use $(N_{\rm pot}, N_{\rm traj}) = (1,100)$.
    }
    \label{fig: EE}
  \end{center}
\end{figure}

\begin{figure}[htb]
  \begin{center}
    \includegraphics[width=\linewidth]{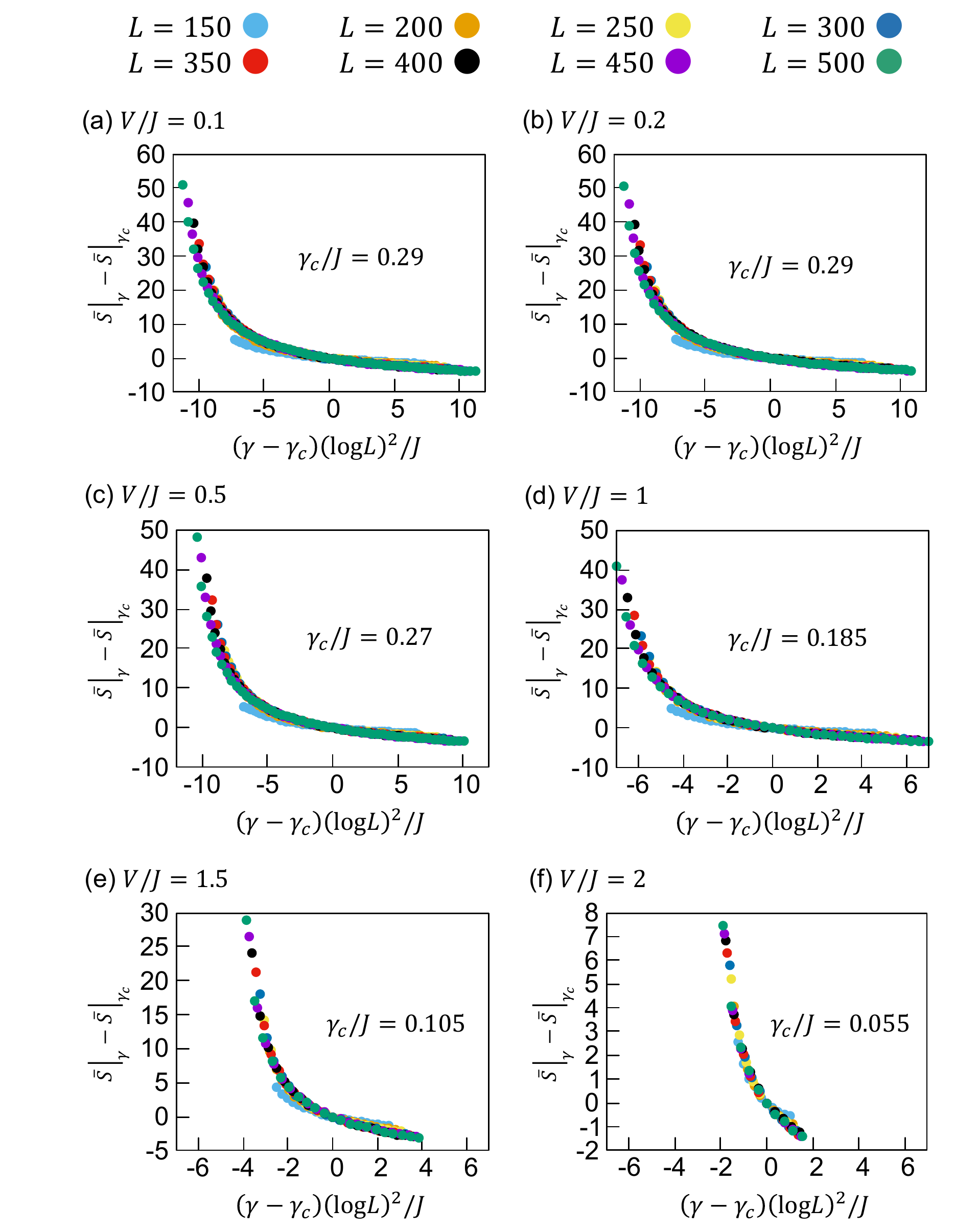}
    \caption{
    Finite-size scaling analysis of the entanglement entropy
    in the systems with $L=150, 200, \cdots, 500$
    when (a) $V/J = 0.1$, (b) 0.2, (c) 0.5, (d) 1, (e) 1.5, and \cred{(f) 2}.
    In this simulation, we use $(N_{\rm pot}, N_{\rm traj}) = (1,100)$.
    }
    \label{fig: scaling}
  \end{center}
\end{figure}

Now, we study the system size dependence of the entanglement entropy in detail. 
For $V/J=0.5$, the entanglement entropy logarithmically grows with respect to the system size for $\gamma<\gamma_c(\sim 0.3J)$ 
and the logarithmic-law phase is stabilized, as shown in Fig.~\ref{fig: EE}(a).
On the other hand, for $\gamma>\gamma_c$, the entanglement entropy does not grow, 
indicating that the area-law phase II is realized.
This fact means the existence of the MIPT between these two phases.
Here, we note the difference of the system size dependence of the entanglement entropy between the volume-law phase ($\gamma/J=0$) and the logarithmic-law phase $(\gamma/J=0.01$).
When $L$ is small, only a tiny difference appears in entanglement entropy, as shown in Fig.~\ref{fig: EE}(a). 
As $L$ increases, the ratio $\overline S |_{\gamma /J=0.01}/\overline S |_{\gamma /J=0}$ tends to decrease monotonically;
for example, the ratio is approximately $0.7$ for $L=500$.
This suggests that continuous monitoring immediately induces the logarithmic-law phase.

As $V$ increases, the magnitude of the entanglement entropy decreases as a whole,
while the qualitative behavior does not change until $V/J\le 2$,
as shown in Figs.~\ref{fig: EE}(a) and \ref{fig: EE}(b).
However, a further increase  of $V$ leads to an intriguing behavior.
The results for $V/J=2.2$ are shown in Fig.~\ref{fig: EE}(c).
In the unitary limit ($\gamma/J=0$), the system is in the area-law phase I and 
the entanglement entropy is around $\bar S\sim3$ irrespective of the system size.
However, we observe logarithmic growth in the entanglement entropy for $\gamma/J=0.01$.
This indicates that continuous monitoring immediately induces the logarithmic-law phase
rather than the area-law phase.
An important point is that 
the logarithmic-law phase persists up to $V_c\:(\sim 2.3J)$.
This phase boundary differs from that in the unitary limit,
where the property of wave functions drastically changes at $V/J=2$.
This behavior is attributed to the destructive effect of measurements on the localized wave function~\cite{Szyniszewski_2023}. 
As $\gamma$ increases further, the entanglement entropy becomes nearly independent of the system size 
and the area-law phase II is realized again, as shown in Fig.~\ref{fig: EE}(c).

When the potential strength is large enough, 
we find no growth of the entanglement entropy irrespective of the measurement strength
although the data for weak $\gamma$ slightly
exceeds the value without measurement, as shown in Fig.~\ref{fig: EE}(d).
This indicates the absence of MIPT in this region.

\subsection{Finite-size scaling analysis}
\label{sec: scaling}

Here, we determine the phase transition point more precisely.
This is because estimating it using the relation $c_{\rm eff}=0$ is hard
due to the finite size effect,
which has been pointed out in previous studies~\cite{Alberton_2021, Minato_2022, Muller_2022, Szyniszewski_2023}.
The entanglement entropy should be scaled around 
the transition point between logarithmic-law phase and area-law phase II as, 
\begin{equation}
  \left. \overline S  \right|_{\gamma} -
  \left. \overline S  \right|_{\gamma_c}
  = F[ (\gamma - \gamma_c) (\log L)^2 ],
  \label{eq: scaling}
\end{equation} 
where $F$ is the scaling function.
This scaling formula is based on the assumption that 
the MIPT belongs to
the universality class similar to a BKT one~\cite{Harada_1997},
where the correlation length $\xi$ diverges exponentially around the transition point; 
$\log \xi \sim 1 / \sqrt{\gamma-\gamma_c}$ as $\gamma \rightarrow \gamma_c +0$.

We use the finite-size scaling analysis for the cases with several $V/J$.
In Fig.~\ref{fig: scaling}, 
we find that the appropriate values of $\gamma_c$
give fairly good scaling plots.
This is consistent with the scaling formula \eqref{eq: scaling}.
To determine the transition point and its error bar, 
we minimize the cost function~\cite{Szyniszewski_2023}, 
where the error bar of $\gamma_c$ is estimated from the range within twice of its minimum value.
The obtained results are shown as the circles with error bars in Fig.~\ref{fig: phase-diagram}, and summarized in Table~\ref{table}.
We find that the critical measurement strength $\gamma_c$ decreases monotonically as $V$ increases.
This result is different from the non-monotonic phase boundary reported in disordered free fermions~\cite{Szyniszewski_2023}. 
This may lead to measurement-induced properties unique to quasiperiodic systems 
that are not seen in monitored dynamics under uniform disorder.
We have also tested the finite-size scaling with $
  \left. \overline S  \right|_{\gamma} -
  \left. \overline S  \right|_{\gamma_c}
  = F[ (\gamma - \gamma_c) (\log L) ]$, and confirmed that the data do not change the results qualitatively, although it may affect the quantitative difference of the transition points.

\begin{table}[h]
  \centering
  \caption{Transition point $\gamma_c$ and its error determined by the finite-size scaling analysis. }
  \begin{tabular}{c|cc}
  \hline
  \hline
  Potential strength $V/J$ & Transition point $\gamma_c / J$ & Error \\ \hline
  0.1 &0.29 &$\pm$0.1 \\ 
  0.2 & 0.29 & $\pm$0.1 \\ 
  0.5 & 0.27 & $\pm$0.1 \\ 
  1 & 0.185 & $\pm$0.065 \\ 
  1.5 & 0.105 & $\pm$0.035\\
  \cred{2} & \cred{0.055} & \cred{$\pm$0.015}\\ \hline\hline
  \end{tabular}
  \label{table}
\end{table}

\section{Further results to support the MIPT}\label{sec: support}

To support the existence of the MIPT
in the 1D free fermion systems under the quasiperiodic potential,
we also examine the connected correlation functions,
single-particle wave functions,
and autocorrelation functions.

\subsection{Connected correlation functions}

\begin{figure*}[htb]
  \begin{center}
    \includegraphics[width=\linewidth]{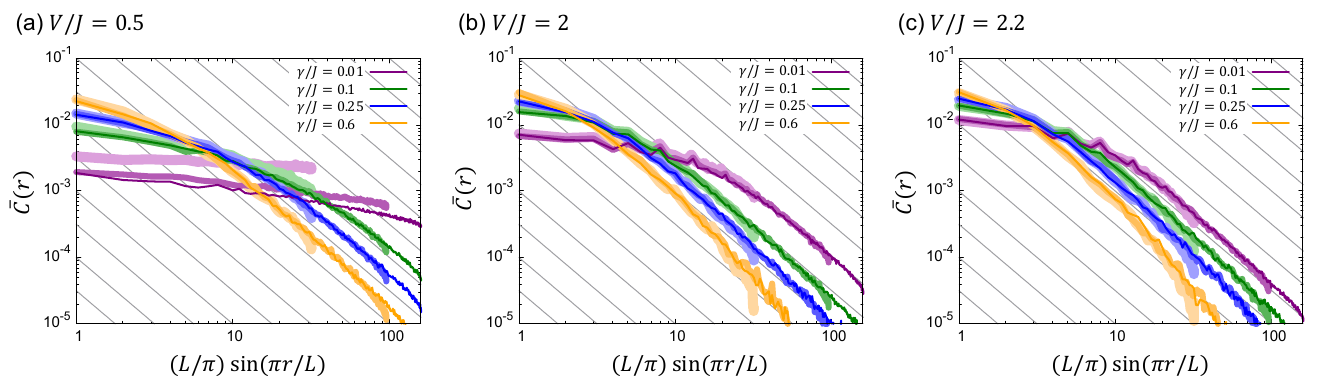}
    \caption{Connected correlation function $\overline{C}(r)$
    for the potential strength (a) $V/J = 0.5$, (b) 2, and (c) 2.2.
    The curves with light, medium, and dark colors
    represent the results for system sizes $L = 100$, 300, and 500,
    respectively.
    Gray lines show the decay $\propto	 [(L/\pi) \sin (\pi r /L)]^{-2}$,
    which approaches $r^{-2}$ in the thermodynamic limit $L \rightarrow \infty$.
    In this simulation, we use $(N_{\rm pot}, N_{\rm traj}) = (46, 23)$.
    }
    \label{fig: CF}
  \end{center}
\end{figure*}

\begin{figure}[htb]
  \begin{center}
    \includegraphics[width=0.7\linewidth]{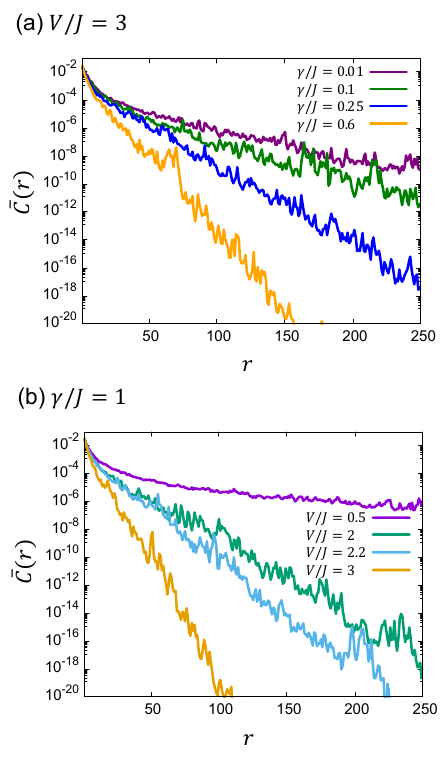}
    \caption{Connected correlation function $\overline{C}(r)$
    in the area-law phase II
    for (a) the potential strength $V/J = 3$, and
    (b) the measurement strength $\gamma / J = 1$.
    In this simulation, we use $L=500$ and $(N_{\rm pot}, N_{\rm traj}) = (46,23)$.
    }
    \label{fig: CF_area}
  \end{center}
\end{figure}

The connected correlation function is defined by
\begin{equation}
C(r) = \langle n_j n_{j+r} \rangle -\langle n_j \rangle \langle n_{j+r} \rangle,
\end{equation}
where $\langle\cdot\rangle$ = $\bra{\Psi} \cdot \ket{\Psi}$ is
the expectation value with respect to the quantum state $\ket{\Psi}$,
and $r$ is a distance between two sites.
In the simulation, we treat the systems with $L=100, 300$, and $500$ and calculate $C(r)$ for $r\le L/2$.
The results for several $V/J$ are shown in Fig.~\ref{fig: CF}.
Except for the case of $V/J = 0.5$ and $\gamma/J = 0.01$,
the data collapse onto a single curve,
suggesting that the converged curves represent
those in the thermodynamic limit.
We find that, 
$\overline C(r) \propto [(L/\pi) \sin (\pi r /L)]^{-2}$ in the logarithmic-law phase,
indicating that the correlation function decays algebraically as $\overline C(r) \propto r^{-2}$ in the thermodynamic limit $L \to \infty$.
The power-law decay of connected correlation function has been also reported in previous studies on monitored free fermions \cite{Alberton_2021, Szyniszewski_2023}.
On the other hand, in the area-law phase II,
such as in the case with $\gamma/J=0.6$,
correlation functions tend to decay rapidly, as shown in Fig.~\ref{fig: CF}.
To clarify its distance dependence in the area-law phase II,
we present the logarithmic plot in Fig.~\ref{fig: CF_area}.
We clearly find the exponential decay in the correlation function,
which is in contrast to the power-law decay in the logarithmic-law phase.
We note that, although the curves for $V/J=0.5$ at $\gamma/J = 0.01$ in Fig.~\ref{fig: CF}(a) terminate before exhibiting $1/r^2$ scaling due to numerical limitation, we expect that larger system sizes would reveal the same algebraic decay as other data. 
Therefore, we expect that $C(r)\propto r^{-2}$ in the large $r$ region. 
The above results clarify that the correlation functions exhibit the
distinct behavior between the logarithmic-law phase and area-law phase II.

\subsection{Wave functions}\label{sec: wave}

\begin{figure}[htb]
  \begin{center}
    \includegraphics[width=0.8\linewidth]{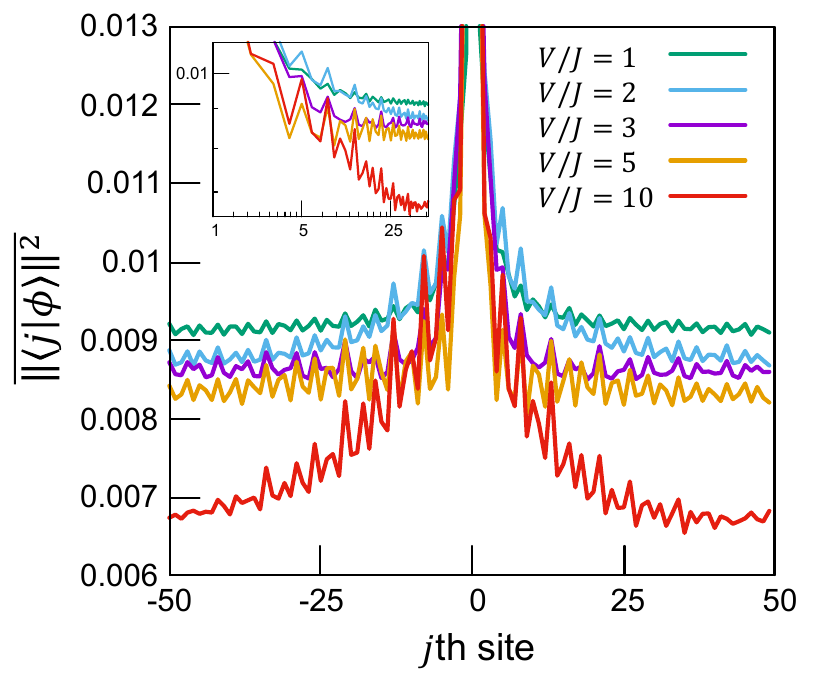}
    \caption{
    Amplitudes of single-particle wave functions in the real space,
    where the inset shows the log-log plot.
    In this simulation, we use $L = 100$, $\gamma / J = 0.01$
    and $(N_{\rm pot}, N_{\rm traj}) = (10^3,5\times 10^2)$.
    }
    \label{fig: single}
  \end{center}
\end{figure}

We examine the localization properties of single-particle wave functions.
We recall that, in the unitary limit $\gamma/J=0$,
the eigenstate wave function is extended (localized) when $V/J<2$ ($V/J>2$),
and critical at $V/J=2$.
Accordingly, distinct single-particle properties in the long-time limit
have been observed in Fig.~\ref{fig: unitary_single}.
Figure~\ref{fig: single} shows the \cred{single-particle} wave function
under the continuous monitoring,
where
we shift the amplitudes in the spatial direction such that their maximum values are positioned at $j=0$. The amplitudes are then averaged over multiple trajectories, resulting in the clear behavior of the localized wave functions.
When $V/J=1$,
the single-particle wave function is similar to that
observed in the unitary limit
since a finite amplitude exists at each site.
This may imply that
this quantity is not appropriate for distinguishing
the volume-law phase in the unitary limit and the logarithmic-law phase under measurement.
However, a drastic change appears for $2\leq V/J \leq 5$.
In particular, the wave function away from the center of the system
is changed by the continuous measurement;
its amplitude is zero when $\gamma/J=0$,
while it is finite when $\gamma/J=0.01$.
Therefore, the single-particle wave function may capture
the measurement-induced logarithmic-law phase.
When $V/J=10$,
the localization of the wave function becomes more pronounced,
as shown in the inset of Fig.~\ref{fig: single}.
This indicates that the system is in the area-law phase II
characterized by the localized wave functions.
\cred{Furthermore, as given in Appendix~\ref{app: IPR}, the inverse participation ratio (IPR) quantitatively supports the validity of this discussion.}
Although determining the phase transition point remains challenging,
this analysis provides insight into these phases.

\subsection{Autocorrelation functions}

\begin{figure}[htb]
  \begin{center}
    \includegraphics[width=0.8\linewidth]{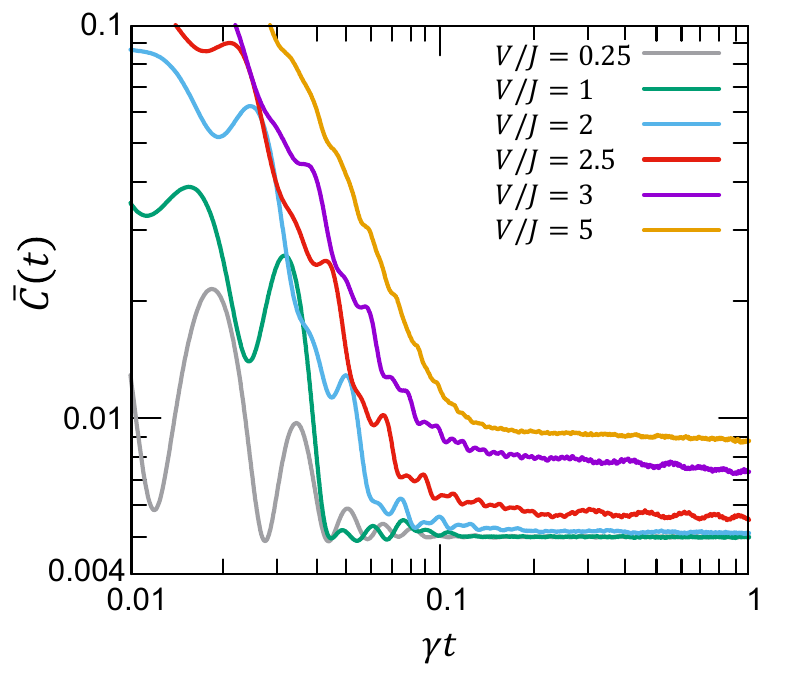}
    \caption{Autocorrelation functions $\overline{C} (t)$.
    The systems are set for $L = 100$ and $\gamma / J = 0.01$.
    In this simulation, we use $(N_{\rm pot}, N_{\rm traj}) = (1,10^6)$.
    }
    \label{fig: AC}
  \end{center}
\end{figure}

Finally,
we examine the autocorrelation function at $j$th site, which is given by
\begin{equation}
  C(t) = \langle n_{j,t^\prime + t} n_{j,t^\prime} \rangle -  \langle n_{j,t^\prime + t} \rangle \langle n_{j,t^\prime} \rangle,
\end{equation}
where $t$ is the time measured from the time $t'$, the latter of which is the time required to reach the steady state.
This quantifies the correlations of the system between different times.
The results for the $\gamma/J=0.01$ case are shown in Fig.~\ref{fig: AC}.
The autocorrelation function does not converge to a constant value within the finite simulation time $\gamma t_{\rm max}=1$,
but exhibits intriguing transient behavior, as shown in Fig.~\ref{fig: AC}.
For $\gamma t \lesssim 0.1$,
the quantity rapidly decays with oscillations.
To further analyze the behavior in the region with slower decay, we focus on the quantity at the time $\gamma t=1$.
For $V/J < 2$, $\overline{C}(t)$ remains approximately $5 \times 10^{-3}$
with a slight increase as the potential strength increases.
On the other hand, for $V/J > 2$, $\overline{C}(t)$ increases significantly with increasing the potential strength, 
as shown in Fig.~\ref{fig: AC}.
Such a rapid increase of the autocorrelation function means that the monitored quantum state 
is unlikely to change due to localization effects, 
which is consistent with the results for the properties of the entanglement entropy 
shown in the phase diagram [see Fig.~\ref{fig: phase-diagram}(a)].

\section{Summary}\label{sec: summary}
We have studied the dynamics under continuous measurement
for free fermions in a quasiperiodic potential by using the AAH model.
We have revealed the existence of the robust logarithmic phase for weak potential strength.
It has been found that
a further increase in the measurement strength induces the MIPT to the area-law phase. 
For strong potential strength, we have demonstrated that MIPTs vanish and the area-law phase is stable irrespective of the measurement strength. 
We have further analyzed connected correlation functions and single-particle wave functions 
to support the above results: 
connected correlation functions exhibit an algebraic decay and 
amplitudes of single-particle wave functions tend to a constant at the edges in the logarithmic-law phase. 
These findings highlight the distinct behavior of MIPTs in quasiperiodic systems compared to the unitary limit and 
lead to the understanding of entanglement dynamics in quasiperiodic quantum systems.
\cred{We note that the destruction of localized or insulating states under dephasing has been studied in the context of dephasing-enhanced transport~\cite{Arenas_2013,Znidaric_2017,Znidaric_2013,Ferreira_2024}, which has been also observed in quasiperiodic systems \cite{Lacerda_2021}, where weak (strong) dephasing delocalizes (localizes) the wave function when the potential strength is large enough. This is similar to the behavior of the wave function for $2< V/J < 2.3$ in the AAH model under continuous measurement, and investigating the effect of quasiperiodicity on the dynamics described by Lindblad equations is also intresting.}

\begin{acknowledgments}
  Parts of the numerical calculations are performed in the supercomputing systems in ISSP, the University of Tokyo.
  T.M. was supported by JST SPRING, Japan Grant Numbers JPMJSP2106 and JPMJSP2180.
  This work was supported
  by Grant-in-Aid for Scientific Research from JSPS, KAKENHI Grant Numbers JP23K19031, JP25K17327 (K.Y.), and JP22K03525 (A.K.). K.Y. was also supported by Murata Science and Education Foundation, Hirose Foundation, the Precise Measurement Technology Promotion Foundation, and the Fujikura Foundation.
\end{acknowledgments}

\section*{Data availability}
The data that support the findings of this article are openly available~\cite{zenodo}.

\appendix

\begin{figure*}[htb]
  \begin{center}
    \includegraphics[width=\linewidth]{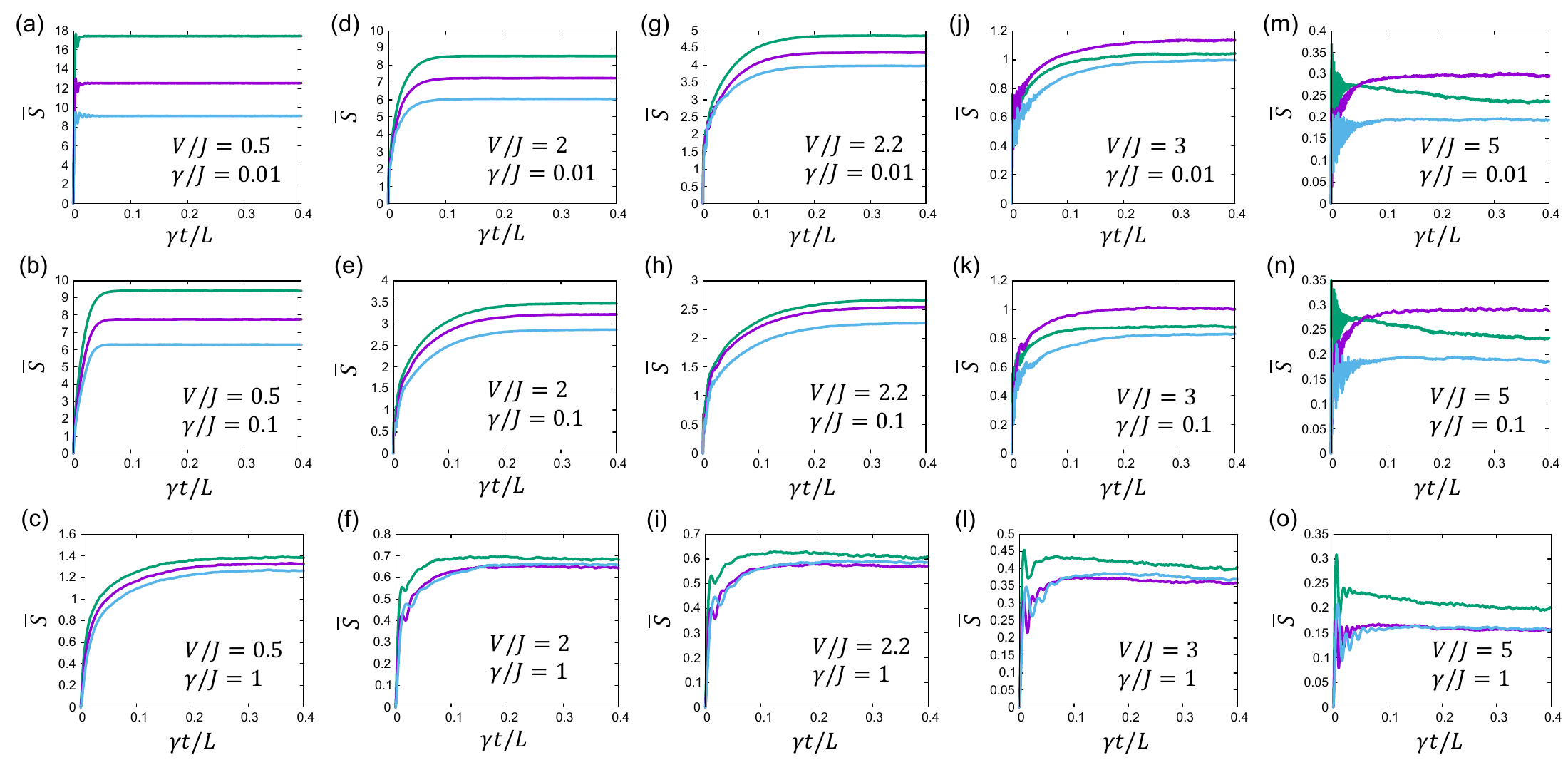}
    \caption{
    Dynamics under continuous measurement for the averaged entanglement entropy by changing the potential strength $ V $ and the measurement strength $ \gamma $.
    The blue, purple, and green lines correspond to the system sizes $ L = 50$, $ 70 $, and $ 100 $, respectively.
    In this simulation, we use $(N_{\rm pot}, N_{\rm traj}) = (1,10^4)$
    }
    \label{fig: converge}
  \end{center}
\end{figure*}

\section{Method for the numerical simulation}\label{app: method}
Here, we provide the details of the numerical simulation for the trajectory evolution described by Eq.~\eqref{eq: QJ}. In general, there is a restriction of the system size for the numerical simulation of the measurement-induced dynamics because we need to employ an exact diagonalization of the system Hamiltonian and follow the long-time evolution. However, in our study, we can simulate the dynamics since the wave function for quantum trajectories preserves the Gaussian state \cite{Alberton_2021, Muller_2022, Minato_2022, Szyniszewski_2023}, even in the presence of the quasiperiodic potential.

For each trajectory, the quantum state at time $ t $ is represented as a Gaussian state, which is
parametrized by an $ L \times N_f $ matrix $ U(t) $ as follows:
\begin{equation}
  \left|\Psi_t\right\rangle = \prod_{k=1}^{N_f}\left(\sum_{j} U_{j, k}(t) c_j^{\dagger}\right)|{\rm vac}\rangle,
\end{equation}
where $ N_f $ is the number of particles, and $ U^\dagger (t) U (t) = I_{N_f} $ is satisfied for the identity matrix $I_{N_f}$ because the wave function is normalized after each quantum jumps. We note that $ \left|\Psi_t\right\rangle $ should reflect the fermionic statistics, where the single-particle wave function that is constructed from $k$-th column of the matrix $U (t)$ is expressed as
\begin{equation}
\left|\phi_k(t)\right\rangle = \sum_j U_{j, k}(t)|j\rangle.
\end{equation}
Here, $ |j\rangle $ represents a wave function that is localized at the $ j $-th lattice site.
The preservation of the Gaussian state structure follows from the fact that
Eq.~\eqref{eq: QJ} represents the time evolution described by a quadratic fermionic operator.
Specifically, the time evolution of the wave function is given by
\begin{equation}
U(t+dt) \propto e^m e^{-ihdt} U(t),
\end{equation}
where the elements of the matrices $h$ and $m$ are defined as
\begin{eqnarray}
h_{i,j} &=& -J(\delta_{i,j+1} + \delta_{i,j-1}) + V\cos (2\pi i/\tau +\theta)\delta_{i,j},\\
m_{i,j} &=& \xi_{i, t} \left( \frac{n_i}{\sqrt{\left\langle n_i \right\rangle_t}} - 1 \right)\delta_{i,j},
\end{eqnarray}
where $\delta_{i,j}$ is the Kronecker delta.

The essence of this calculation is that we can easily compute physical quantities by using the so-called correlation matrix $ D (t, t^\prime)$, which is given by
\begin{equation}
  D_{j,k}\left(t, t^{\prime}\right) = \left[U\left(t\right) U^{\dagger}(t^{\prime})\right]_{j,k} = \left\langle c_{j, t}^{\dagger} c_{k, t^{\prime}}\right\rangle.
\end{equation}
To simulate the quantum-jump evolution in Eq.~\eqref{eq: QJ}, we focus on the fact that the particle number
is conserved even in the presence of quantum jumps described by $n_i$. Then, the quantum jumps occur at time $t$, which does not depend on the quantum state, and this fact is different from the quantum trajectory evolution for the non-Hermitian quantum jump operator \cite{Daley_2014, Yamamoto_2022}. The procedure is outlined in detail in Ref.~\cite{Alberton_2021}, but we briefly summarize the algorithm in Table~\ref{algorithm} for the sake of readability.
\begin{table*}[hbt]
  \centering
  \caption{Algorithm to calculate the matrix $U (t)$ for the Gaussian state by using the quantum trajectory method with the particle number measurement.}
  \begin{tabular}{l}
  \hline
  \hline
  1. Determine the quantum jump time $ \tau = -\log(p) / \gamma N_f $, where $ p $ is a random number uniformly distributed in the interval $ [0, 1] $. \\
  $\quad$ Next, evolve the system to time $ t + \tau $ by using the equation $ U(t + \tau) = e^{-iH\tau} U(t) $. \\ \\

  2. Select a jump operator $ n_j $ based on the probability $ P(n_j) = \braket{n_j}_{t+\tau}/N_f $. \\ 
  $\quad$ Then, update the correlation matrix $ D(t+\tau, t+\tau) = U(t + \tau)U^\dagger(t + \tau) $ according to:\\ \\
  $\quad$
  $
    D^\prime_{l m} (t+\tau, t+\tau) =
    \begin{cases} 
      \delta_{l,j}\delta_{m,j}, & (l = j \text{ or } m = j), \\ \\
      D_{l m}(t+\tau, t+\tau) -  {D_{j m} (t+\tau, t+\tau) D_{l j} (t+\tau, t+\tau)}/
      \braket{n_j}_{t+\tau},
      & \text{(otherwise)},
    \end{cases}
  $ 
  \\ \\ $\quad$ where $D^\prime (t+\tau, t+\tau)$ is the correlation matrix after the quantum jump.\\ \\
  3. Reconstruct the updated matrix $ U^\prime (t+\tau) $ by performing an SVD decomposition of the Hermitian matrix
  $ D^\prime (t+\tau)$,\\ $\quad$ where
$D^\prime (t+\tau) = U^\prime (t+\tau) \Sigma U^{\prime\dagger} (t+\tau)$, $\quad \Sigma_{11} = \ldots = \Sigma_{N_f N_f} = 1$, and $\quad \Sigma_{N_f +1,N_f +1} = \ldots = \Sigma_{LL} = 0$. \\ \\
  4. Repeat the processes 1-3 by choosing another random number. \\\hline\hline
  \end{tabular}
  \label{algorithm}
\end{table*}

\section{Physical observables}\label{app: obs}
Here, we provide the formulas used to compute physical observables. 
We assume that the correlation matrix $ D(t, t^\prime) $ is obtained for a fermionic chain of length $ L $. First, the von Neumann entanglement entropy $ S $ for a subsystem $ A = [m_1, m_2] $ of length $ \ell = |m_2 - m_1 + 1| $
is calculated from the eigenvalues $ \{\lambda_j^{(\mathrm{A})}\} $ of the reduced equal-time correlation matrix $ D^{(\mathrm{A})}(t, t) $, which is defined as $D^{(\mathrm{A})}(t, t) = D_{j=m_1, \ldots, m_2, k=m_1, \ldots, m_2}(t, t)$~\cite{Calabrese_2005, Alba_2018}.
Then, the entanglement entropy at time $t$ is given by
\begin{equation}
  S = -\sum_{j=1}^\ell \left[ \lambda_j^{(\mathrm{A})} \log \lambda_j^{(\mathrm{A})} 
  + \left(1 - \lambda_j^{(\mathrm{A})}\right) \log \left(1 - \lambda_j^{(\mathrm{A})}\right) \right].
\end{equation}
Similarly, we obtain the connected correlation function $ C(r) $ from the correlation matrix as
\begin{equation}
  C(r) = | D_{j+r,j} (t,t) |^2.
\end{equation}
The autocorrelation function $ C(t) $ is calculated as
\begin{equation}
  C(t) = | D_{j,j}(t^\prime, t^\prime+t) |^2.
\end{equation}

\section{Dynamics of the entanglement entropy under continuous measurement}\label{app: steady}

Here, we analyze the dynamics of the entanglement entropy before saturating to the steady state. As shown in Fig.~\ref{fig: converge}, the time required for the system to reach the seady-state value of the entanglement entropy is roughly proportional to the system size $ L $ and inversely proportional to the measurement strength $ \gamma $ when the potential strength is fixed.
On the other hand, there exists a nontrivial dependence on the potential strength $ V $. We find in Figs.~\ref{fig: converge}(a)-(o) that we need to evolve the state longer to let the system reach the steady state as $ V $ is increased.
Moreover, we find that the time evolution of the entanglement entropy reveals distinct behavior depending on the potential strength $V$.
For the potential strength $V/J = 0.5, 2, 2.2,$ and $3$ [see Figs.~\ref{fig: converge}(a)-(k)],
the entanglement entropy grows almost monotonically over time regardless of the measurement strength. We note that, in Figs.~\ref{fig: converge}(l) for $V/J = 3$ and $\gamma/J=1$, the entanglement entropy starts to oscillate and shows nonmonotonic behavior before saturation. This is caused by the coexistence of the localization effect originating from the quasiperiodic potential and the measurement. Such effects of the quasiperiodic potential is much more enhanced when the potential strength is increased for $V/J=5$ [see Figs.~\ref{fig: converge}(m)-(o)]. We clearly find that the entanglement entropy displays nonmonotonic growth, exhibiting strong oscillations before reaching the steady state. To study such nontrivial oscillations is interesting and deserves further study.

\section{Inverse participation ratio}\label{app: IPR}

\cred{To quantitatively characterize the localization of the wave function,
we employ IPR as one of the standard measures.
The IPR is defined as 
\begin{equation}
I = \sum_{j=0}^{L-1} ||\braket{j|\phi}||^4,
\end{equation}
where $\ket{j}$ is the state localized at the $j$th site, and $\ket{\phi}$ is the single-particle wave function that we evaluate.  For a state localized at a single site (e.g., $\ket{\phi}=\ket{0}$), IPR takes the maximum value as $I=1$.  On the other hand, for an extended state such as $|\braket{j|\phi}|^2=1/L$, the IPR takes the minimum value as $I=1/L$.  Therefore, if IPR becomes large, it signifies stronger localization of the wave function. }

\begin{figure}[htb]
  \begin{center}
    \includegraphics[width=0.8\linewidth]{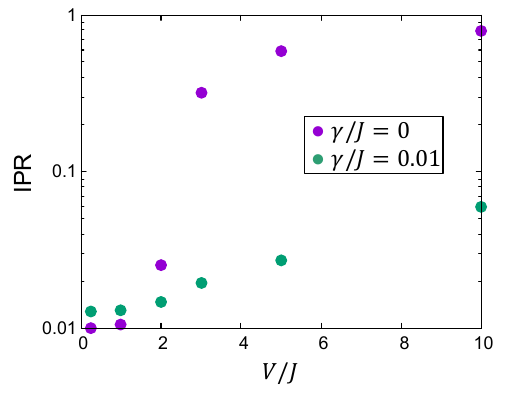}
    \caption{\cred{The IPR of the single-particle wave function. The calculation is performed for $L=100$, where the minimum value of IPR is $I=0.01$.}
    }
    \label{fig: IPR}
  \end{center}
\end{figure}

\cred{We now calculate IPR to quantitatively evaluate the localization strength of the wave functions shown in Figs.~\ref{fig: unitary_single} and \ref{fig: single}. The results are presented in Fig.~\ref{fig: IPR}. In the unitary limit ($\gamma/J=0$), IPR increases sharply when the potential strength exceeds $V/J=2$.  This result is consistent with the fact that the wave function becomes localized above $V/J=2$.  For $2 \le V/J \le 5$ with the measurement strength $\gamma/J=0.01$, we see that IPR is suppressed compared to the unitary limit.  This result provides quantitative support for the fact that the wave function away from the center is finite under measurement, while it becomes zero in the unitary limit.  Furthermore, we find that IPR is enhanced for $\gamma/J=0.01$ as the potential strength is increased.  This is consistent with our discussion that the wave function becomes more localized for strong potential strength such as $V/J=10$. These results quantitatively characterize the localization behavior shown in Sec.~\ref{sec: wave}.}

\nocite{apsrev42Control}
\bibliographystyle{apsrev4-2}
\bibliography{./refs}

\begin{thebibliography}{127}
\bibitem [{com()}]{comment}%
  \BibitemOpen
  \href@noop {} {}\bibinfo {note} {As discussed in Ref.~\cite{Jian_2020}, we
  should not confuse $c_\mathrm{eff}$ with the true central charge that
  reflects the universality of the model \cite{Calabrese_2004, Calabrese_2009},
  because $c_\mathrm{eff}$ includes the contribution from the exponent of
  boundary correlation functions in subsystem $A$. Therefore, $c_\mathrm{eff}$
  can be variant depending on the measurement strength.}\BibitemShut {Stop}%
\end{thebibliography}

\end{document}